\documentclass[journal]{IEEEtran}
\usepackage{latexsym,,amssymb,amsmath,graphicx,epsf,cite,bbm,float}
\usepackage{ifpdf}
\usepackage{epstopdf,mathtools}

\usepackage{algorithm,algorithmic}
\usepackage{amsmath,amssymb,bm}
\usepackage{amsfonts,dsfont,color,bbm,subcaption}

\usepackage{mathtools}

\def\ninept{\def\baselinestretch{1}}
\ninept

\newcommand{\abs}[1]{|#1|}

\DeclareMathOperator*{\argmax}{arg\,max}
\DeclareMathOperator*{\argmin}{arg\,min}

\usepackage{hyperref,amsthm}

\newtheorem{theorem}{Theorem}

\newtheorem{lemma}[]{Lemma}

\newtheorem{proposition}[]{Proposition}

\newtheorem{remark}[]{Remark}

\newtheorem{definition}[]{Definition}

\newtheorem{assumption}[]{Assumption}

\begin{document}

\title{
	A Linearithmic Time Locally Optimal Algorithm for the Multiway Number Partition Optimization
} 
\author{\IEEEauthorblockN{Kaan Gokcesu}, \IEEEauthorblockN{Hakan Gokcesu} }
\maketitle

\begin{abstract}
	We study the problem of multiway number partition optimization, which has a myriad of applications in the decision, learning and optimization literature. Even though the original multiway partitioning problem is NP-hard and requires exponential time complexity algorithms; we formulate an easier optimization problem, where our goal is to find a solution that is locally optimal. We propose a linearithmic time complexity $O(N\log N)$ algorithm that can produce such a locally optimal solution. Our method is robust against the input and requires neither positive nor integer inputs.
\end{abstract}

\section{Introduction}\label{sec:intro}
\subsection{Multiway Number Partition Problem}

The multiway number partitioning ($K$-way number partitioning) \cite{graham1969bounds,mertens2006number,cook1971complexity,turing1937computable,levin1973universal,korf1998complete} is the problem of partitioning a set of numbers $\mathcal{X}$ into $K$ number of subsets $\{\mathcal{X}_k\}_{k=1}^K$ such that the individual sums $\{S_k\}_{k=1}^K$ of the subsets $\{\mathcal{X}_k\}_{k=1}^K$ are as similar as possible.
In this work, we deal with the optimization version of the multiway number partition problem. The exact optimization objective can be defined in a number of ways such as the maximization of the minimum set sum, the minimization of the maximum set sum or the minimization of the difference between the maximum and the minimum set sums; all of which are equivalent when $K=2$, but they are all different when $K\geq3$ \cite{korf2010objective,walter2013comparing}.

A closely related problem is the number partition problem, where the partition is done over $K=2$ subsets \cite{korf1998complete}.
Another closely related problem is the subset-sum problem, where the goal is to find a subset of a set, whose sum equals a target value $T$ \cite{kleinberg2006algorithm}. One more closely related problem is the bin packing problem, where the goal is to find a partition with the smallest possible number of subsets $K$ given that the subset sum is bounded \cite{hochbaum1987using}.
Unfortunately, all of these problems are NP-complete for their combinatorial versions and NP-hard for their optimization versions \cite{karp1972reducibility,garey1979computers,korf2009multi}.

Despite the hardness, for the number partition problems, there exist efficient methods to solve them in many instances. There can be an exponential number of optimal solutions, which makes one possibly easier to identity. All in all, because of its limited structure, number partitioning a comparatively easier problem than other NP-hard problems  \cite{hayes2002computing,mertens2006number}.

Even though, it is an NP-hard problem and requires exponential in $N$ (number of samples) time complexity, there exists polynomial time approximation methods and pseudo-polynomial time (dependent on the input values) exact algorithms \cite{garey1979computers}. 
It is a well studied subject with an ongoing extensive research and all of these approaches can be promising given the appropriate application (even though pseudo-polynomial algorithms may have limited uses for high precision inputs) \cite{graham1966bounds,coffman1978application,dell1995optimal,korf1998complete,korf2009multi,moffitt2013search,schreiber2013improved,schreiber2014cached,schreiber2018optimal}.

\subsection{Applications of Multiway Number Partitioning}\label{sec:application}
The multiway number partition problem has a lot of applications in learning, optimization and decision problems, e.g., scheduling, encryption, allocation, selection, dataset partitioning \cite{neyshabouri2018asymptotically,cano2006combination,cano2007evolutionary,gokcesu2018adaptive,garcia2009enhancing,kellerer2004knapsack,golbraikh2000predictive,fan2005working,huberbook,cesabook,gokcesu2020generalized,poor_book,furusjo2006importance,gokcesu2021generalized,sarkar1987partitioning,mathews1896partition,gokcesu2020recursive,dell2008heuristic,graham1979optimization,gokcesu2021optimal,walsh2009really,merkle1978hiding,shamir1982polynomial,rivest1983cryptographic,gokcesu2021optimally,buhrkal2011models,umang2013exact,iris2015integrated,lalla2016set,gokcesu2021regret,friesen1981analysis,deuermeyer1982scheduling,coffman1984approximation,dantzig2007number,brams1996fair,biswas2018fair,johnson1991optimization,biggs1991method,loh1997split,kim2001classification,kim2004memory,gokcesu2021efficient,dasdan1997two,kose2020novel,liu1998network,gokcesu2022low,chatha2001magellan,jiang2007temporal,karthikeyan2019key}. Given the optimization objective, some prominent examples are as follows.

\subsubsection{Minimize the difference between max and min set sums}
The first popular objective is the minimization of the difference between the largest set sum and the smallest set sum. This is a common objective in research about multiway number partitioning \cite{mertens2006number}. An example application is the problem of choosing fair teams \cite{hayes2002computing}. Let each element $x$ in the input set $\mathcal{X}$ correspond to a player's ability and the power of a team be equal to the sum of the player abilities. The objective is to create $K$ teams, where the strongest and the weakest teams are as close as possible to each other.

\subsubsection{Minimize the maximum set sum}
Another popular objective is the minimization of the largest set sum. This objective is commonly called the processor scheduling problem in some literature \cite{garey1979computers,sarkar1987partitioning,dell2008heuristic,graham1979optimization}. The input set $\mathcal{X}$ of $N$ positive elements correspond to the individual run-times of a set of $N$ tasks. The aim here is to assign each task to one of the identical machines (such as parallel processor cores) such that we minimize the total time it takes to complete all the tasks, i.e., the last task should be completed as early as possible.
Similarly, the berth allocation problem \cite{buhrkal2011models,umang2013exact} in the field of operations research deals with the allocation of berth space in container terminals for incoming vessels. Here, the operator needs to assign arriving vessels to berths for container loading (or unloading) such that they are taken care of as soon as possible. This is a multiway partitioning problem given the number of berths $K$ and incoming vessels loading (or unloading) times $\mathcal{X}$ \cite{iris2015integrated,lalla2016set}.

\subsubsection{Maximize the minimum set sum}
One last popular objective is the maximization of the minimum set sum. This objective commonly arises in fair division \cite{brams1996fair,biswas2018fair}. Moreover, it appears in sequencing maintenance actions for modular engines \cite{friesen1981analysis,deuermeyer1982scheduling}. Let us have $K$ number of engines that we want to keep alive, where each engine needs a certain critical part for its operation. Let us have $N$ number of that part with possibly different lifespan $\mathcal{X}$. This is equivalent to maximizing the minimum set sum when we want to keep the engine with the shortest lifetime as long as possible. 
Another example is the case of veto election, where voters veto a candidate (each veto has a different weight) \cite{walsh2009really}. If the candidate with the smallest total veto wins, a group's best strategy will be to partition their veto weights among the opposing candidates and maximizing the minimum set sum.

\subsection{Algorithms in Literature} 
Given an input set $\mathcal{X}$ of size $N$, the most straightforward algorithm for the multiway partition problem is the brute force approach, which has $O(K^N)$ time complexity. The problem is NP-hard and its solution takes exponential time to find. 
There exist efficient sub-optimal approaches.
Most notable ones are the greedy number partitioning method \cite{graham1966bounds}, the multifit algorithm \cite{coffman1978application} and the largest differencing method (i.e., Karmarkar-Karp set differencing algorithm) \cite{karmarkar1982differencing}, which can run in $O(N\log N)$ time and $O(N)$ space \cite{kellerer2004knapsack,korf2011hybrid}.
There also exists approximate methods, whose runtime complexities are polynomial in the number of elements $N$ and exponential in the approximation parameter $\epsilon^{-1}$ \cite{graham1969bounds,hochbaum1987using,sahni1976algorithms,woeginger1997polynomial,alon1998approximation}
There are dynamic programming approaches \cite{garey1979computers,martello1990knapsack,korf2009multi,korf2013optimally}, which can find an optimal solution in pseudo-polynomial time and space, where the complexities are polynomially dependent on the maximum of the input set $\mathcal{X}$. Note that, in such approaches the inputs are assumed to be positive and integer. Thus, their performance is highly dependent on the precision of the inputs. 
The complete anytime algorithm in \cite{korf1998complete} can use sub-optimal algorithms as decision heuristics to generate optimal algorithms. It creates a $K$-ary tree from the sub-optimal algorithm selections (much like a modification of the brute force approach) to create anytime algorithms with linear memory usage. However, its worst-case time complexity is exponential.
There are also hybrid algorithms \cite{schreiber2018optimal}, which combines the complete anytime algorithm and other methods from the subset sum problem and the bin packing problem to achieve an even better performance (albeit still exponential in the worst case).
There are also algorithms which can produce locally optimal solutions for the set partition problem when $K=2$ in polynomial time \cite{gokcesu2021efficient, gokcesu2021quadratic}. Instead of a seemingly arbitrary sub-optimal heuristic, a locally optimal solver may prove to be more useful in many scenarios. 

\subsection{Contributions and Organization}
Although the sub-optimal algorithms have polynomial runtime, their solutions can be significantly far from an optimal. While approximate methods' runtime is polynomial in $N$, they are exponential in the approximation precision $\epsilon$. The pseudo-polynomial algorithm is an exact solver but has limited use for high precision or non-integer inputs. Even though, the complete anytime algorithm is an exact solver with better runtime than brute force, its complexity is still exponential in the worst case. Improving the exponential complexity is futile because of the NP-hardness of the problem but fast algorithms are always desired especially with the emergence of big data. Although \cite{gokcesu2021efficient} design efficient algorithms for a locally optimal solution, they are only applicable when $K=2$.
To this end, we tackle the 'weaker' version of the multiway number partition problem and extend the results of \cite{gokcesu2021efficient} to the generic case of $K\geq 2$. In \autoref{sec:prob}, we mathematically formulate the problem definition. In \autoref{sec:method}, we provide an efficient algorithm that can find a locally optimal solution in $O(N\log N)$ time and $O(N)$ space. In \autoref{sec:ext}, we extend our methods to any real inputs and finish with concluding remarks.

\section{Locally Optimal \texorpdfstring{$K$}{K}-way Partition Problem}\label{sec:prob}
In this section, we formally define the $K$-way partitioning problem as in \cite{gokcesu2021efficient}. As an input, we have the set of numbers
	\begin{align}
	\mathcal{X}=&\{x_1,x_2,\ldots,x_N\},\\
	=&\{x_n\}_{n=1}^N.
	\end{align}	
We partition $\mathcal{X}$ into $K$ subsets $\{\mathcal{X}_k\}_{k=1}^K$ (for some natural number $K\geq 2$, which differs from \cite{gokcesu2021efficient}) such that they are disjoint and their union is $\mathcal{X}$, i.e.,
	\begin{align}
	\mathcal{X}_{i}\cap\mathcal{X}_j&=\emptyset, &&\forall {i,j}\in\{1,\ldots,K\}; i\neq j\\
	\cup_{k=1}^K\mathcal{X}_k&=\mathcal{X},
	\end{align}	
The goal is to create the sets $\{\mathcal{X}_k\}_{k=1}^K$ such that their individual sums are as close as possible to each other. Hence, in this problem, we compare the set sums with each other, which are denoted as
\begin{align}
S_k=\sum_{x\in\mathcal{X}_k}x,&& k\in\{1,\ldots,K\}
\end{align}
Hence, $\sum_{k=1}^{K}S_k=S$, where
\begin{align}
	S=\sum_{x\in\mathcal{X}}x.
\end{align}
There are different schools of though for the formulation of the problem. As given in \autoref{sec:application}, some examples are:
\begin{itemize}
	\item Minimization of the maximum set sum: $$\min_{\{\mathcal{X}_k\}_{k=1}^K}\max_k S_k$$
	\item Maximization of the minimum set sum: $$\max_{\{\mathcal{X}_k\}_{k=1}^K}\min_k S_k$$
	\item Minimization of the maximum difference:
	$$\min_{\{\mathcal{X}_k\}_{k=1}^K}(\max_k S_k-\min_k S_k)$$
\end{itemize}
In the case of $K=2$, i.e., two-way partitioning, all such formulation are equivalent since we are dealing with only two sets and their respective sums. However, when $K\geq 3$, this is not the case; and different formulations produces different optimization problems.

These problems are unfortunately NP-hard \cite{karp1972reducibility,gokcesu2021efficient,gokcesu2021quadratic}, and is impossible to solve with an efficient method. To this end, instead of these NP-hard problem, we consider a 'weaker' version \cite{gokcesu2021efficient} of the $K$-way partitioning problem. Instead of a global optimal solution, we are after a 'locally' optimal one, which is defined as follows:

\begin{definition}\label{def:local}
	A $K$-way partitioning $\{\mathcal{X}_k\}_{k=1}^K$ is locally optimal if there is no single element transfer that can decrease the absolute difference between the sums of its former and latter sets, i.e.,
	\begin{align}
		|(S_i-x)-(S_j+x)|\geq |S_i-S_j|, \forall x\in\mathcal{X}_i \text{ and }\forall {i,j}
	\end{align}
\end{definition}

The local optimality definition in \autoref{def:local} is universal in the sense that such a solution is locally optimal for all of the different formulations mentioned before.
Next, we provide an efficient procedure which can find such a locally optimal $K$-way partitioning.

\section{A Linearithmic Complexity Method}\label{sec:method}
\subsection{Iterative Algorithm}\label{sec:alg}
Before we propose the algorithm, we make some initial assumptions similar to \cite{gokcesu2021efficient}.
\begin{assumption}
	Let the set $\mathcal{X}$ be composed of only positive elements, i.e.,
	\begin{align*}
		x>0, &&\forall x\in \mathcal{X}.
	\end{align*}
\end{assumption}
\begin{assumption}
	Let the set $\mathcal{X}=\{x_n\}_{n=1}^N$ be in ascending order, i.e.,
	\begin{align*}
	x_n\leq x_{n+1}, &&\forall n\in\{1,2,\ldots,N-1\}.
	\end{align*}
\end{assumption}
Thus, our input $\mathcal{X}$ is a positive ordered set. 

\begin{remark}
	If the set is not ordered, we can do a simple merge sort to sort the set in $O(N\log N)$ time and $O(N)$ space \cite{skiena1998algorithm}. 
\end{remark}

\begin{remark}
	We point out that there is no requirement for the elements to be integers.
\end{remark}

Given the input $\mathcal{X}$ set, the algorithm works as follows: 
\begin{enumerate}
	\item Put all elements into the first set, i.e., $\mathcal{X}_1\equiv \mathcal{X}$ and $\mathcal{X}_k=\emptyset$ for $k\geq 2$. Thus, $S_1=S$, where $S$ is the sum of all elements in $\mathcal{X}$ and $S_k=0$ for $k\geq 2$. Create the index set $\mathcal{K}=\{1,\ldots,K\}$
	
	\item If $\abs{\mathcal{K}}=1$, STOP;\\
	else continue.\label{step:ite}

	\item Find an index $i$ of a set with the largest sum, i.e., $i=\argmax_k S_k$.
	
	\item Find an index $j$ of a set with the smallest sum, i.e., $j=\argmin_k S_k$.
	
	\item Let $\mathcal{X_*}$ be the set of elements whose move decreases the absolute difference between the set sums $S_i$ and $S_j$, i.e.,
	\begin{align}
		\mathcal{X_*}=\{x\in\mathcal{X}_i: \abs{(S_i-x)-(S_j+x)}<\abs{S_i-S_j}\}
	\end{align}\label{step:absdiff}

	\item IF $|\mathcal{X}_*|=0$ (i.e., empty), set $\mathcal{K}\leftarrow \mathcal{K}\setminus\{i\}$, return to Step \ref{step:ite};\\
	ELSE set $x_*=\max_{x\in\mathcal{X_*}}x$.
	
	\item Move $x_*$ from $\mathcal{X}_i$ to $\mathcal{X}_j$ and update the sets and their sums accordingly. Return to Step \ref{step:ite}.
\end{enumerate}

\begin{remark}
	A few remarks about the algorithm:
	\begin{itemize}
		\item At the beginning of our algorithm, all elements of $\mathcal{X}$ are assigned to a single set $\mathcal{X}_1$, i.e.,
		\begin{align}
			\mathcal{X}_1\equiv\mathcal{X}, && \mathcal{X}_k\equiv\emptyset, \forall k\neq 1.
		\end{align}
		\item At the beginning, we have the set sums 
		\begin{align}
			S_1=\sum_{x\in\mathcal{X}}x, &&S_k=0, \forall k\neq 1.
		\end{align} 
		\item At each iteration of the algorithm (from Step \ref{step:ite} to itself), we either discard a maximum sum set $\mathcal{X}_i$, or move a single element from a maximum sum set $\mathcal{X}_i$ to a minimum sum set $\mathcal{X}_j$. 
	\end{itemize}
\end{remark}

We start our analysis of the algorithm with some important observations.
 
\subsection{Preliminaries}

We start with some preliminary results and show that the algorithm definitely terminates as in \cite{gokcesu2021efficient}.
\begin{proposition}\label{thm:S_ijcloser}
	After each element move between sets $\mathcal{X}_i$ and $\mathcal{X}_j$, we have the following at Step \ref{step:ite}:
	\begin{align*}
		\max(S_i^{new},S_j^{new})<&\max(S_i^{old},S_j^{old}),\\
		\min(S_i^{new},S_j^{new})>&\min(S_i^{old},S_j^{old}),
	\end{align*}
	where $S_i^{old}, S_j^{old}$ and $S_i^{new},S_j^{new}$ are the set sums before and after the element move respectively.
	\begin{proof}
		Let the set sums be $S_i^{old}$ and $S_j^{old}$ before moving an element $x\in\mathcal{X}_i$. We have
		\begin{align}
			S_i^{old}=\frac{1}{2}(\mu^{old}+\delta^{old}),&&
			S_j^{old}=\frac{1}{2}(\mu^{old}-\delta^{old}),
		\end{align}
		where $\mu^{old}=S_i^{old}+S_j^{old}$ and $\delta^{old}=S_i^{old}-S_j^{old}$. Let 
		\begin{align}
			S_i^{new}=\frac{1}{2}(\mu^{new}+\delta^{new}),&&
			S_j^{new}=\frac{1}{2}(\mu^{new}-\delta^{new}),
		\end{align}
		for some $\mu^{new}$ and $\delta^{new}$, similarly. Since
		\begin{align}
			S_i^{new}=S_i^{old}-x,&&S_j^{new}=S_j^{old}+x
		\end{align}
		for some $x\in\mathcal{X}_i$, we have $\mu^{new}=\mu^{old}$. Moreover, we know from Step \ref{step:absdiff} that the absolute difference strictly decreases, hence, $\abs{\delta^{new}}<\delta^{old}$. Thus, we have 
		\begin{align}
			\mu^{old}-\delta^{old}<\mu^{new}-\abs{\delta^{new}}\leq \mu^{new}+\abs{\delta^{new}}< \mu^{old}+\delta^{old},
		\end{align}
		which concludes the proof.
	\end{proof}
\end{proposition}

\begin{proposition}\label{thm:S_kcloser}
	After each element move, we have
	\begin{align*}
		\max_k S_k^{new}\leq \max_k S_k^{old},\\
		\min_k S_k^{new}\geq \min_k S_k^{old},
	\end{align*}
	where $\{S_k^{old}\}_{k=1}^K$ and $\{S_k^{new}\}_{k=1}^K$ are the set sums before and after the move respectively. 
	\begin{proof}
		When an element $x$ is moved from set $\mathcal{X}_i$ to $\mathcal{X}_j$; it does not change the set sums $S_k$, where $k\notin\{ i,j\}$. From \autoref{thm:S_ijcloser}, we know that $\max (S_i,S_j)$ decreases and $\min (S_i,S_j)$ increases, which concludes the proof.
	\end{proof}
\end{proposition}

\begin{proposition}\label{thm:S_kcloserstrictly}
	The values of the maximum set sum and the minimum set sum strictly decreases and increases respectively after a finite number of element transfers in the algorithm.
	\begin{proof}
		From \autoref{thm:S_ijcloser}, we know that when a transfer happens between two set sums, their maximum and minimum strictly decreases and increases respectively. Since there is a finite number of sets, the minimum and the maximum set sums strictly decreases and increases respectively after a finite number of iterations.
	\end{proof}
\end{proposition}

\begin{lemma}\label{thm:terminate}
	The algorithm definitely terminates.
	\begin{proof}
		From \autoref{thm:S_kcloserstrictly}, we know that the maximum and the minimum set sums strictly decreases and increases respectively after a finite number of iterations. Since there is a finite number of elements $N$, there exists a finite number of possible maximum and minimum set sum values. Hence, after a finite number of iterations, the algorithm terminates.
	\end{proof}
\end{lemma}

\subsection{Local Optimality}\label{sec:localopt}
In this section, we prove the local optimality of our algorithm as in \cite{gokcesu2021efficient}. We start with a few useful results and build the local optimality claim from there.

\begin{proposition}\label{thm:x>x_min}
	If moving the element $x_*\in\mathcal{X}_i$ from $\mathcal{X}_i$ to $\mathcal{X}_j$ cannot decrease the absolute distance, neither can any $x\in\mathcal{X}:x\geq x^*$.
	\begin{proof}
		The proof comes from the fact that if moving $x^*$ cannot decrease the absolute difference, it means 
		\begin{align}
			x^*\geq \abs{S_i-S_j}.
		\end{align}
		For any $x\geq x^*$, we have
		\begin{align}
			x\geq \abs{S_i-S_j}.
		\end{align}
		which concludes the proof.
	\end{proof}
\end{proposition}
\begin{proposition}\label{thm:X1XK}
	Let us have a set of sets $\{\mathcal{\tilde{X}}_k\}_{k=1}^K$ whose sums are nonincreasing, i.e., $\tilde{S}_k\geq \tilde{S}_{k+1}$. If there is no element in $\mathcal{\tilde{X}}_1$ whose move to $\mathcal{\tilde{X}}_K$ can decrease $\abs{\tilde{S}_1-\tilde{S}_K}$; there exists no element in $\mathcal{\tilde{X}}_1$ whose move to $\mathcal{\tilde{X}}_k$ can decrease $\abs{\tilde{S}_1-\tilde{S}_k}$ for all $k$.
	\begin{proof}
		Since $\tilde{S}_1$ is maximum, $\mathcal{\tilde{X}}_1$ cannot be empty. Since there exist no element whose move can decrease $\abs{\tilde{S}_1-\tilde{S}_K}$, we have
		\begin{align}
			x\geq {\tilde{S}_1-\tilde{S}_K}, &&\forall x\in\mathcal{\tilde{X}}_1.
		\end{align}
		Since $\tilde{S}_K$ is less than or equal to all $\tilde{S}_k$, we have
		\begin{align}
			x\geq {\tilde{S}_1-\tilde{S}_k}, &&\forall x\in\mathcal{\tilde{X}}_1; \forall k,
		\end{align} 
		which concludes the proof.
	\end{proof}
\end{proposition}

\begin{lemma}\label{thm:Xmax}
	At some point in the algorithm, let $\mathcal{X}_i$ and $\mathcal{X}_j$ be sets with maximum and minimum sum respectively, i.e., $S_i\geq S_k\geq S_j, \forall k$. If there exists no $x\in\mathcal{X}_i$ whose move to $\mathcal{X}_j$ can decrease $\abs{S_i-S_j}$; there can never be an $x\in\mathcal{X}_i$ whose move to $\mathcal{X}_k$ can decrease $\abs{S_i-S_k}$ for all $k$ for the duration of the algorithm.
	\begin{proof}
		From \autoref{thm:S_kcloser}, we know that the maximum and the minimum set sums are nonincreasing and nondecreasing respectively. Thus, if there exists no $x\in\mathcal{X}_i$ whose move to the set with minimum sum can decrease the absolute set difference, there can never be. Moreover, from \autoref{thm:X1XK}, if moving to the set with minimum sum does not decrease the absolute difference, we cannot decrease the absolute difference with any set. Henceforth, if a maximum sum set cannot get closer to a minimum sum set at any point, it can never get closer to any set since it will remain a maximum sum set from \autoref{thm:S_kcloser}, which concludes the proof.
	\end{proof}
\end{lemma}

\begin{theorem}
	The algorithm terminates at a locally optimal $K$-way partitioning solution.
	\begin{proof}
		From \autoref{thm:Xmax}, we know that if a maximum sum set cannot get closer to a minimum sum set at any point, it satisfies local optimality. The algorithm continues by discarding such maximum sum sets until only a single set remains. Hence, at the end, all the sets will satisfy the local optimality criterion, which concludes the proof.
	\end{proof}
\end{theorem}

\subsection{Complexity Analysis}\label{sec:complexity}

Here, we prove the linearithmic time complexity of our method as in \cite{gokcesu2021efficient}.
\begin{proposition}\label{thm:leaveorder}
	While $\mathcal{X}_1$ stays a maximum sum set, the elements leave $\mathcal{X}_1$ in a nonincreasingly ordered fashion.
	\begin{proof}
		We know from \autoref{thm:S_kcloser} that the maximum and the minimum set sums are nonincreasing and nondecreasing respectively. From Step \ref{step:absdiff}, the maximum valued feasible element is moved. Hence, a larger element cannot be transfered after a smaller element, which concludes the proof.
	\end{proof}
\end{proposition}

\begin{proposition}\label{thm:arriveorder}
	While $\mathcal{X}_1$ stays a maximum sum set, the elements arrive at $\mathcal{X}_i$ for any $i$ in a nonincreasing fashion.
	\begin{proof}
		The proof follows \autoref{thm:leaveorder}.
	\end{proof}
\end{proposition}

\begin{lemma}\label{thm:locOpt}
	If at some point in the algorithm, $\mathcal{X}_1$ stayed a maximum sum set but there is no element left that can decrease the absolute sum difference, we have a local optimal solution.
	\begin{proof}
		From \autoref{thm:S_kcloser}, we know that the maximum and the minimum set sums are nonincreasing and nondecreasing respectively. If $\mathcal{X}_1$ cannot get closer to a minimum sum set at some point, it can never do so, hence, it is local optimal. Let us look at the next largest sum set $\mathcal{X}_i$ for some $i$, whose minimum valued element is $x_n$ for some $n$. From \autoref{thm:arriveorder}, we know that $x_n$ is the last arrived element in $\mathcal{X}_i$. From the algorithm, we know that $\mathcal{X}_i$ can only receive $x_n$, when it is a minimum sum set. After receiving $x_n$, the set sum of $\mathcal{X}_i$ is at most $x_n$ greater that the minimum set sum, which cannot decrease from \autoref{thm:S_kcloser}. Hence, moving $x_n$ can never make $\mathcal{X}_i$ closer to a minimum sum set, which means no element in $\mathcal{X}_i$ can ever make it closer to a minimum sum set from \autoref{thm:x>x_min}. Following \autoref{thm:X1XK}, $\mathcal{X}_i$ cannot get closer to any set, which makes it locally optimal. Following a similar argument, all sets are locally optimal, which concludes the proof.   
	\end{proof}
\end{lemma}

\begin{lemma}\label{thm:X1greater}
	If at some point in the algorithm, $\mathcal{X}_1$ is no longer a maximum sum set, the sets with greater or equal sums are locally optimal.
	\begin{proof}
		The proof follows from a similar argument as the proof of \autoref{thm:locOpt}. From \autoref{thm:arriveorder}, a maximum sum set is at a distance at most its minimum element from the minimum set sum. Hence, it is locally optimal. Similarly, all sets with greater or equal set sums are locally optimal.
	\end{proof}
\end{lemma}

\begin{lemma}\label{thm:X1moves}
	In our algorithm, only $\mathcal{X}_1$'s elements are moved.
	\begin{proof}
		The proof follows from \autoref{thm:X1greater}, which implies an element that is already moved from $\mathcal{X}_1$ to another set cannot be moved a second time.
	\end{proof}
\end{lemma}

\begin{theorem}
	The algorithm has $O(N)$ time complexity.
	\begin{proof}
		The proof follows from \autoref{thm:X1moves}. Since every element is moved at most once from $\mathcal{X}_1$, there are at most $O(N)$ moves. Since these moves are in nonincreasing order, we can check feasible moves in descending order. Each check and movement takes $O(1)$ time. Moreover, updating the ordered set sums takes at most $O(\log K)$ time. Hence, we have $O(N\log N)$ time complexity.
	\end{proof}
\end{theorem}

\section{Discussions and Conclusion}\label{sec:ext}
We observe that if $\mathcal{X}$ contains $0$ valued elements, it is inconsequential since they can be arbitrarily assigned to $\mathcal{X}_k$ for any $k$ at the end \cite{gokcesu2021efficient}. Secondly, if the sample set includes not only positive samples but also negative ones, we can deal with them with the following changes to the algorithm:

\begin{enumerate}
	\item Put all positive elements into the first set and all negative elements into the last set, i.e., $\mathcal{X}_1=\{x\in\mathcal{X}:x>0\}$, $\mathcal{X}_K=\{x\in\mathcal{X}:x<0\}$ and $\mathcal{X}_k=\emptyset$ for $2\leq k \leq K-1$. Thus, $S_1=S_+$ and $S_K=S_-$, where $S_+$ and $S_-$ are the sum of all positive and negative elements respectively; and $S_k=0$ for $2\leq k\leq K-1$. Create the index set $\mathcal{K}=\{1,\ldots,K\}$
	
	\item If $\abs{\mathcal{K}}\leq1$, STOP;\\
	else continue.\label{step:iteN}

	\item Find an index $i$ of a set with the largest sum, i.e., $i=\argmax_k S_k$.
	
	\item Find an index $j$ of a set with the smallest sum, i.e., $j=\argmin_k S_k$.
	
	\item Let $\mathcal{X_+}$ and $\mathcal{X_-}$ be the sets of positive and negative elements respectively whose move decreases the absolute difference between the set sums $S_i$ and $S_j$, i.e.,
	\begin{align}
		\mathcal{X_+}=&\{x\in\mathcal{X}_i: \abs{(S_i-{x})-(S_j+{x})}<\abs{S_i-S_j}\}\\
		\mathcal{X_-}=&\{x\in\mathcal{X}_j: \abs{(S_i+{x})-(S_j-\abs{x})}<\abs{S_i-S_j}\}
	\end{align}\label{step:absdiffN}

	\item If $\abs{\mathcal{X_+}\cup\mathcal{X_-}}=0$,set $\mathcal{K}\leftarrow \mathcal{K}\setminus\{i,j\}$ return to Step \ref{step:ite};
	\\
	ELSE set $x_*=\argmax_{x\in\mathcal{X_+}\cup\mathcal{X_-}}\abs{x}$.
	
	\item Move $x_*$ between $\mathcal{X}_i$ and $\mathcal{X}_j$, then update the sets and their sums accordingly. Return to Step \ref{step:ite}.
\end{enumerate}

\begin{remark}
	Similar to \autoref{thm:terminate}, this version of the algorithm also definitely terminates, since the values of the maximum and the minimum set sums strictly changes in finite number of iterations.
\end{remark}

\begin{remark}
	This version of the algorithm also terminates at a locally optimal solution. Its proof follows similar arguments to the ones in \autoref{sec:localopt}. Since locally optimal sets are jointly discarded, the final solution is locally optimal.
\end{remark}

\begin{remark}
	This version of the algorithm takes linearithmic $O(N\log N)$ time to terminate. Its proof follows similar arguments to the ones in \autoref{sec:complexity}. Since every element can only be moved once, we have at most $O(N)$ moves.
\end{remark}

In conclusion, we studied the optimization version of the $K$-way partitioning problem, which is a multiway generalization of the number partition problem \cite{gokcesu2021efficient}. While this set partition problem is NP-hard and requires exponential complexity to solve; we formulated a weaker version of this NP-hard problem, where the goal is to find a locally optimal solution. We proposed an algorithm that can find locally optimal solutions in linearithmic $O(N\log N)$ time and linear $O(N)$ space. Our algorithms require neither positive nor integer elements in the input set, hence, they are widely applicable.

\bibliographystyle{IEEEtran}
\bibliography{double_bib}

\end{document}